\documentclass[aps,prd,twocolumn,nofootinbib,floatfix]{revtex4-1}
\usepackage{setspace}
\usepackage{amsthm}
\theoremstyle{definition}

\usepackage{graphicx}
\usepackage{dcolumn}
\usepackage{multirow}
\usepackage{subfigure}
\usepackage{times,mathptm}
\usepackage{float}
\usepackage{color}
\usepackage{amsmath,amsfonts}
\usepackage{mathptmx}
\usepackage{mathrsfs}
\usepackage{bbm}
\usepackage{bm}
\usepackage{xfrac}

\newcommand{\beq}{\begin{equation}}
\newcommand{\eeq}{\end{equation}} 
\newcommand{\bea}{\begin{eqnarray}}
\newcommand{\eea}{\end{eqnarray}}

\newcommand{\E}{\mathcal{E}}

\newcommand{\pbar}{\overline{\psi}}
\renewcommand{\d}{\delta}

\renewcommand{\b}{\beta}

\newcommand{\vx}{{\vec{x}}}

\newcommand{\tbar}{\overline{\bf 3}}
\newcommand{\tbf}{{\bf 3}}

\newcommand{\e}{\epsilon}

\newcommand{\dg}{\dagger}
\newcommand{\non}{\nonumber}

\newcommand{\rf}[1]{(\ref{#1})}
\newcommand{\ra}{\rightarrow}

\renewcommand{\vec}[1]{\bm #1}

\usepackage{ulem}

\begin{document}

\title{Screening without screening:  baryon energy at high baryon density} 

\bigskip
\bigskip

\author{Jeff Greensite and Evan Owen}
\affiliation{Physics and Astronomy Department, San Francisco State
University,   \\ San Francisco, CA~94132, USA}
\bigskip
\date{\today}
\vspace{60pt}
\begin{abstract}

\singlespacing
 
     We compute the Coulomb interaction energy of dense sets of static quarks in a compact volume (much smaller than
the lattice volume) containing one quark per lattice site.  The quark color charges are combined into either a set of three-quark nucleon states, or into  a non-factorizable ``one big hadron'' state.   In both cases we find that the energy per quark is
roughly constant as the volume of quarks increases.  A surprise is that if we construct the nucleon states
from sets of three quarks chosen at random in the volume, then the energy per quark remains
roughly constant, even as the average distance between quarks in a nucleon state grows as
the volume increases.  This energy dependence of a nucleon in a dense medium is at odds with the behavior of an 
isolated nucleon as quark separation increases, and for static quarks it is not easily explicable in terms
of some version of Debye screening.

\end{abstract}

\maketitle

\singlespacing
\section{\label{intro}Introduction}

    The study of QCD at high baryon density is constrained by the (as yet) unsolved sign problem.  There are, however,
some situations related to QCD at high baryon density where the sign problem is less acute.  In this article we will study the Coulomb interaction energy of a dense system of static quark charges in a fixed volume, and for this problem we will see that standard Monte Carlo methods will suffice.  The thermodynamics of a system of heavy dense quarks at finite chemical potential has already been treated by other (e.g.\ Langevin) methods \cite{Aarts:2016qrv,Langelage:2014vpa}, but our focus here is a little different, and goes a little beyond
the phase diagram.  Rather than introducing a
chemical potential, and having a finite baryon density throughout the lattice volume, we will place $2^p \times 3$ quark charges in a subvolume of the lattice, with a density of one quark per site.  This requires that the color charges are contracted into an overall color singlet combination, and we will consider two types of contractions:
\begin{enumerate}
\item A ``multi-nucleon'' (MN) state.  This consists of division of the quarks into $2^p$ sets of three quarks (not necessarily nearest neighbors), and contraction of the quark charges in each set into a color singlet.  This leaves $2^p$ ``nucleon'' states. 
\item A ``di-quark pyramid'' (DQP) state.  Here we first construct $2^{p-1} \times 3$ diquark states in the $\tbar$ representation, form from these $2^{p-2} \times 3$ sets of states in the $\tbf$ representation, and so on until we arrive at three states in either the $\tbf$  (even $p$) or $\tbar$ (odd $p$) representations, which are finally contracted into a singlet.
\end{enumerate}
The point to notice is that unlike the MN state, a DQP state cannot be factorized into two or more subsets of color singlets.  It is, in a sense, one big hadron, where every quark interacts in some way with every other quark.  

Our focus in this article is on energetics.  We would like to know how the color Coulomb
interaction energy per quark depends on volume, and on how the colors are contracted, and on the average distance (in the MN case)
between quarks in a nucleon.  One important point to note from the
beginning is that the usual mechanisms of charge screening in a plasma are not available here, because the charges
are static, and the positions of the quarks are fixed.

\section{\label{contra} Color contractions}

   In this section we use an upper-lower color index convention to distinguish between indices transforming in the $\tbar$
and $\tbf$ representations, respectively, with indices raised and lowered by complex conjugation.  Thus a gauge
transformation $g(x)$ transforms quark fields as
\bea
          \psi_{a}(x) &\ra& \psi'_{a}(x) = g_a^{~~b}(x) \psi_{b}(x) \non \\
         \pbar^{a}(x) &\ra&  \pbar'^{a}(x) = g^a_{~~b}(x) \pbar^{b}(x) \non \\
          g_a^{~~b}(x) g^c_{~~b}(x) &=& g^b_{~~a}(x) g_b^{~~c}(x) = \d^c_a \ .
\eea

   In a theory with local SU(3) gauge invariance, a set of quark operators $\psi_{a}$  in the equal-times 
combination
\footnote{We ignore spin states and Dirac indices; for the Coulomb energy only color combinations are important.}
\beq
          \e^{abc} \psi_{a}(\vx_1) \psi_{b}(\vx_2) \psi_{c}(\vx_3)
\eeq
is invariant under global, but not local SU(3) gauge transformations.  To form a gauge-invariant combination which, when
applied to the vacuum, would create a physical state, we must in general replace  $\e_{abc}$ by a multi-covariant gluonic operator
\beq
         V^{abc}(\vx_1,\vx_2,\vx_3;A) \psi_{a}(\vx_1) \psi_{b}(\vx_2) \psi_{c}(\vx_3) \ ,
\eeq
which transforms under a local gauge transformation $g$ as
\bea
       & & V^{abc}(\vx_1,\vx_2,\vx_3;A) \ra V'^{abc}(\vx_1,\vx_2,\vx_3;A)  \non \\
       & &  \qquad = g^a_{~~a'}(\vx_1) g^b_{~~b'}(\vx_2) g^c_{~~c'}(\vx_3)
       V^{a'b'c'}(\vx_1,\vx_2,\vx_3;A)  \ .
\eea
This obviously generalizes to any number of quarks (providing that number is divisible by three), i.e.\ the creation operator
for a system of $N$ quarks will have the form
\bea
  & &     V^{a_1 a_2 a_3...a_N}(\vx_1,\vx_2,\vx_3,...,\vx_N;A) \non \\
   & & \qquad \times    \psi_{a_1}(\vx_1) \psi_{a_2}(\vx_2)   \psi_{a_3}(\vx_3) \cdot \cdot \cdot   \psi_{a_N}(\vx_N) \ .
   \non \\
\eea
What is meant by a ``multi-nucleon'' (MN) state is that the operator which creates the state can be factorized into products of gauge invariant operators, each composed of three quarks.  So the operator which creates an MN state of six quarks factorizes into a product of two nucleon operators, e.g.
\bea                                                                                       
\Psi^{MN}  &=& V^{a_1 a_2 a_3 a_4 a_5 a_6}(\vx_1,\vx_2,\vx_3,\vx_4,\vx_5,\vx_6) \psi_{a_1}(\vx_1) \non \\
   &\times&   \psi_{a_2}(\vx_2)  \psi_{a_3}(\vx_3)  
   \psi_{a_4}(\vx_4) \psi_{a_5}(\vx_5)   \psi_{a_6}(\vx_6) ~ \Psi_0
\non \\
 &=& \Bigl\{ V^{a_1 a_2 a_3}(\vx_1,\vx_2,\vx_3)\psi_{a_1}(\vx_1) \psi_{a_2}(\vx_2)  \psi_{a_3}(\vx_3) \Bigr\}  \non \\
  &\times&   \Bigl\{ V^{a_4 a_5 a_6}(\vx_4,\vx_5,\vx_6)  
  \psi_{a_4}(\vx_4) \psi_{a_5}(\vx_5)   \psi_{a_6}(\vx_6)  \Bigl\} ~ \Psi_0 \ , \non \\
\eea
where $\Psi_0$ the vacuum state.  A DQP operator cannot be factorized in this manner.     

    We are interested in comparing the energies of MN and DQP states.  Of course it is difficult to do this in
full generality, not least because the $V$ operators which minimize the energies of such states are unknown.  However,
if we are satisfied with just computing numerically the Coulomb interaction energy of such states, then it is possible
to make such comparisons.  By ``Coulomb energy'' we mean the energy, above the vacuum energy, of a state obtained from a set of quark creation operators in Coulomb gauge acting on the vacuum.  In particular, let $A_k(\vx)$ be a gauge field at some fixed time $t$, and let $g(x;A)$ be the gauge transformation which takes a quark operator at point $x$ into Coulomb gauge. Then we may construct, e.g., a gauge-invariant one-nucleon state as
\bea
       \Psi &=&  \e^{a_1 a_2 a_3} g_{a_1}^{~~b_1}(\vx_1;A) g_{a_2}^{~~b_2}(\vx_2;A) g_{a_3}^{~~b_3}(\vx_3;A) \non \\
          & & \times \psi_{b_1}(\vx_1) \psi_{b_2}(\vx_2)   \psi_{b_3}(\vx_3) \Psi_0 \ .
\eea
If $A$ is already in Coulomb gauge, then $g(\vx;A)$ is the identity operator, and therefore in Coulomb gauge we have simply
\beq
  \Psi = \e^{a_1 a_2 a_3}  \psi_{a_1}(\vx_1) \psi_{a_2}(\vx_2)   \psi_{a_3}(\vx_3) \Psi_0 \ .
\label{pc}
\eeq
We note that Coulomb gauge does not fix the gauge uniquely (even if we ignore the Gribov copy issue), because
the gauge condition is preserved by gauge transformations $g(\vx,t)=g(t)$ which are constant on a time-slice.  The contraction of indices with the Levi-Civita tensor is required in order that $\Psi$ is invariant under this global remnant of the gauge symmetry.   The interaction energy of a state of this kind can only be due to the non-local Coulomb term in the Hamiltonian, since there is no other expression in the Coulomb gauge Hamiltonian which could give rise to an interaction energy between spatially separated quarks.  So we will refer to the energy of state \rf{pc} above the vacuum energy as the ``Coulomb energy'' of a set of three static quarks.

   More generally, let $S^\dg$ be the creation operator for a set of static quarks in Coulomb gauge, invariant under the remnant symmetry, and denote
\beq
          |\Psi_S \rangle = S^\dg |\Psi_0 \rangle \ .
\eeq
We define $\E_S(t)$, on a Euclidean lattice with discretized time, from the vacuum expectation values
\bea
           e^{-\E_S(t)} &\equiv& {\langle S(t+1) S^\dg(0) \rangle \over \langle S(t) S^\dg(0) \rangle }   \non \\
                            &=&  {\langle \Psi_S| e^{-(H-E_0)(t+1)} |\Psi_S \rangle \over \langle \Psi_S |e^{-(H-E_0)t} |\Psi_S \rangle } \ ,
\eea
where by $e^{-H n}$ we mean the $n$-th power of the transfer matrix.  Note that
\bea
          \E^{min}_S = \lim_{t\ra \infty} \E_S(t)
\eea
is the minimum possible energy, above the vacuum energy $E_0$, of a state containing the same set of static quarks as 
$\Psi_S$, in the same spatial positions with the same color contractions.   At the other end of the time scale, in the $t=0$ limit,
\bea
          \E  \equiv \E_S(0)
\eea
may be regarded as a definition of the Coulomb energy of state $\Psi_S$ on a discrete time lattice, and it is energies of this kind that we report below.  We of course compute $\E$ on a periodic lattice at each time slice $t$ and average over time slices, so the quantities to be computed by lattice Monte Carlo are ${\langle S(t+1) S^\dg(t) \rangle}$.

   The integration over static quark fields in the $S(t+1) S^\dg(t)$ operator leaves us with a set of timelike link variables on a 
time slice, with one link variable for each static quark position running between times $t$ and $t+1$, and with indices contracted to form a singlet under the remnant global symmetry.  As an example, for the one nucleon state in \rf{pc}, we have for the Coulomb energy
\begin{widetext}
\bea
        \E = -\log\left\{  { \langle \e^{a_1 a_2 a_3}  \e_{b_1 b_2 b_3} 
                [U_0]_{a_1}^{~~b_1}(\vx_1,t)  [U_0]_{a_2}^{~~b_2}(\vx_2,t) [U_0]_{a_3}^{~~b_3}(\vx_3,t) \rangle \over
                 \e^{a_1 a_2 a_3}  \e_{a_1 a_2 a_3} } \right\} \ .  \non \\
\eea
The generalization to larger sets of quarks is straightforward.  In Coulomb gauge the $V$ operators are simply
tensors, independent of position and gauge field, which contract quark indices into global color singlets, and
the Coulomb energy is computed on the lattice from the correlators
\bea
        \E = -\log\left\{  { \langle V^{a_1 a_2 ... a_N}  V_{b_1 b_2 ... b_N} 
                [U_0]_{a_1}^{~~b_1}(\vx_1,t)  [U_0]_{a_2}^{~~b_2}(\vx_2,t) ... [U_0]_{a_N}^{~~b_N}(\vx_N,t) \rangle \over
                  V^{a_1 a_2 ... a_N} V_{a_1 a_2 ... a_N} } \right\}   \non \\
\eea
\end{widetext}
evaluated in Coulomb gauge.
In the past this method has been used to compute the Coulomb energy of a single quark-antiquark pair as a function of
quark separation \cite{Greensite:2003xf,*Nakagawa:2006fk,Greensite:2014bua}.   Our intention
here is to apply the same technique to a dense system of quarks.

    Consider initially an isolated set of $N = 2^p \times 3$ quarks which are in such close proximity that their interactions can be neglected in comparison to their kinetic energies.  Even so, either the quark indices are contracted to form a singlet, or
else we must in addition consider forming a singlet with the help of constituent gluons.  For now we consider only the former possibility.  In a situation of this sort, we may ask what is the most likely contraction of quark color charges that form the singlet.   The simplest possibility is to form $2^p$ nucleon states, each a color singlet of three quarks.  The number 
$\Omega_{MN}$ of ways of making this multi-nucleon grouping is
\beq
   \Omega_{MN} = {N! \over \left({N\over 3}\right)! (3!)^{N/3}} \ .
\eeq
In this counting, the static quark charges are assumed to occupy different lattice sites, and can be treated as distinct objects.
The number of ways of grouping a set of $N$ distinct objects into $N/3$ distinct bins, each containing 3 objects, is
$N!/(3!)^{N/3}$.  But since the ``bins'' are not distinct, and their order is irrelevant, this number must be corrected by
dividing by $(N/3)!$.

Of course there are a vast number of alternate possibilities.  We will not attempt an exhaustive counting, but rather
concentrate on the DQP arrangement.   This is one particular example of a ``one big hadron'' state, as it cannot be factorized into subsets of color singlets, and it is easy to count the DQP multiplicity.  We first divide $N$ quark charges into $N/2$
subsets of two quarks each, i.e.\ diquarks in representation $\tbar$.  The multiplicity is
\beq
     \Omega_1 =  { N! \over \left({N\over 2}\right)! 2^{N/2} } \ .
\eeq 
The next step is to group the $N/2$ diquarks in $N/4$ subsets of two diquarks each, with multiplicity 
\beq
     \Omega_2 =  { \left({N\over 2}\right)!  \over \left({N\over 4}\right)! 2^{N/4} } \ ,
\eeq 
proceeding in this way $p$ times until there are only three charged objects left.  The total multiplicity is then
\bea
          \Omega_{DQP} &=& \Omega_1 \Omega_2 \cdot \cdot \cdot \Omega_p  \non \\
                                    &=& {N! \over 2^{N-3} 3!} \ .
\eea
Obviously there are far more DQP states than multi-nucleon states.  If the energies of the DQP and multi-nucleon states
are comparable,  then purely on the grounds of multiplicity the system is bound to be in a one big hadron state, although not
necessarily in a DQP state.  But in fact diquark pairings are favored energetically at the perturbative level, and it has been argued that this sort of pairing exists even within nucleons \cite{Anselmino:1992vg,*Lichtenberg:1982jp}. It could be that DQP states are favored energetically in general among one big hadron states, although this is, of course, far from certain.  In any case they constitute a simple alternative to MN states, and the Coulomb energies of both DQP and MN states can be computed numerically, and compared at varying densities and spatial arrangements.

\begin{figure*}[htb]
\centering
\subfigure[~ 6 quarks]  
{   
 \label{q6}
 \includegraphics[scale=0.35]{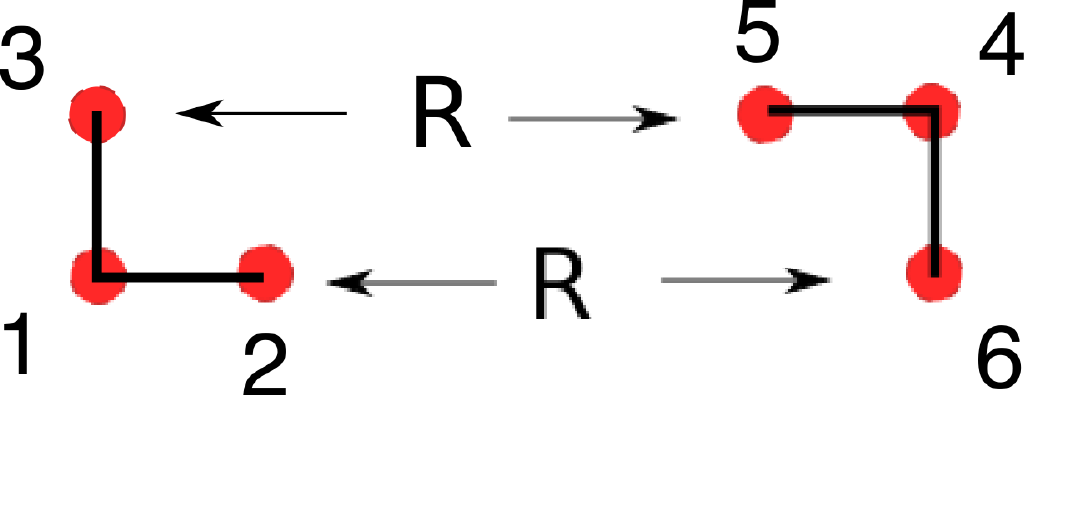}
}
\hfill
\subfigure[~ 12 quarks]  
{   
 \label{q12}
 \includegraphics[scale=0.35]{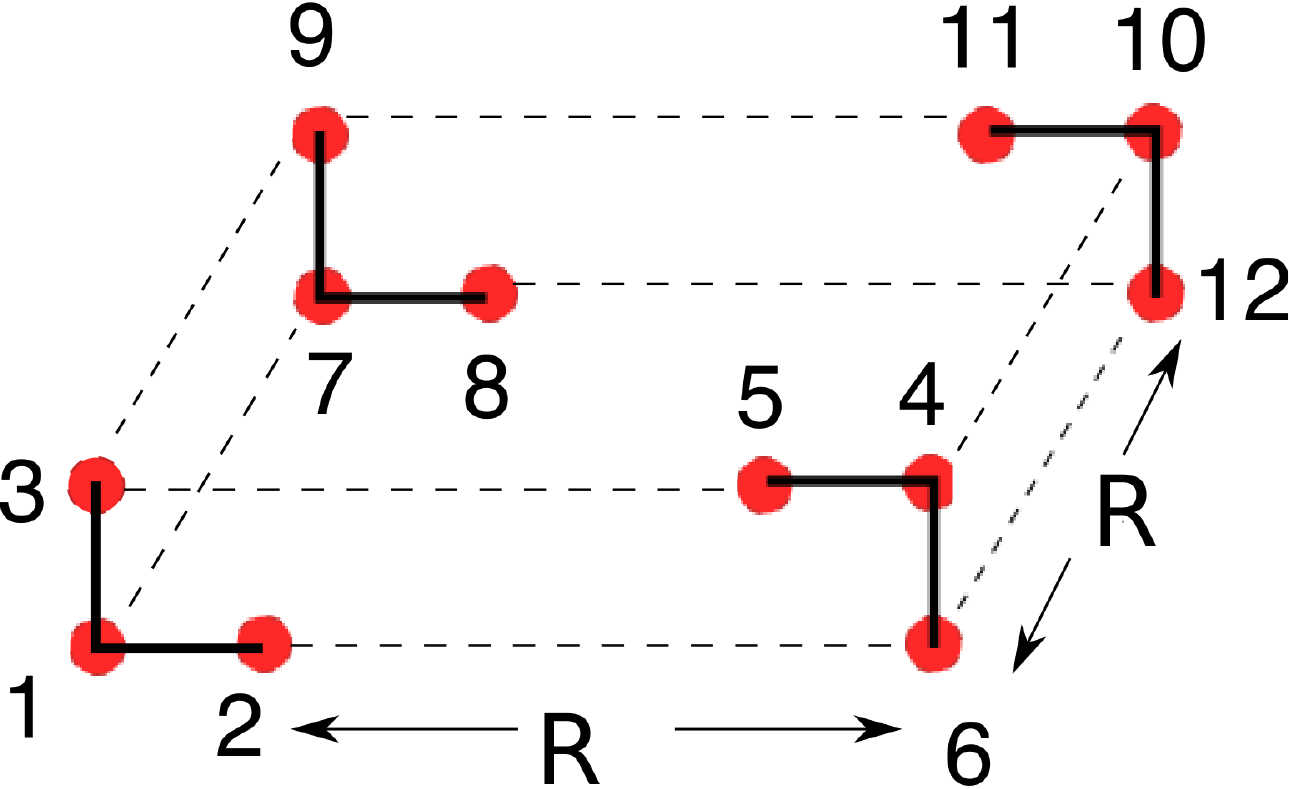}
}
\hfill
\subfigure[~ 24 quarks] 
{   
 \label{q24}
 \includegraphics[scale=0.35]{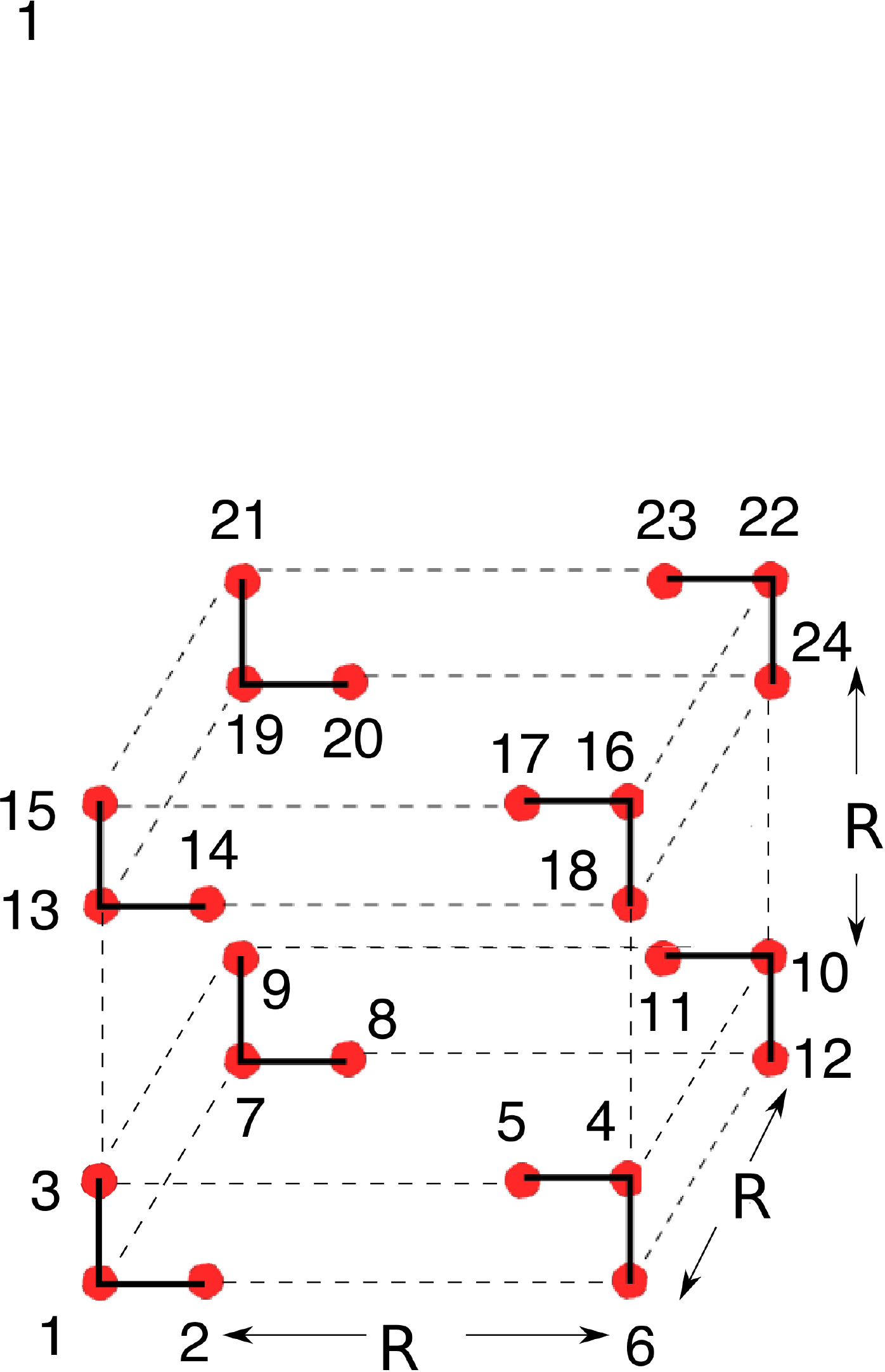}
}
\subfigure[~ 6 quarks]  
{   
 \label{bhgas1}
 \includegraphics[scale=0.25]{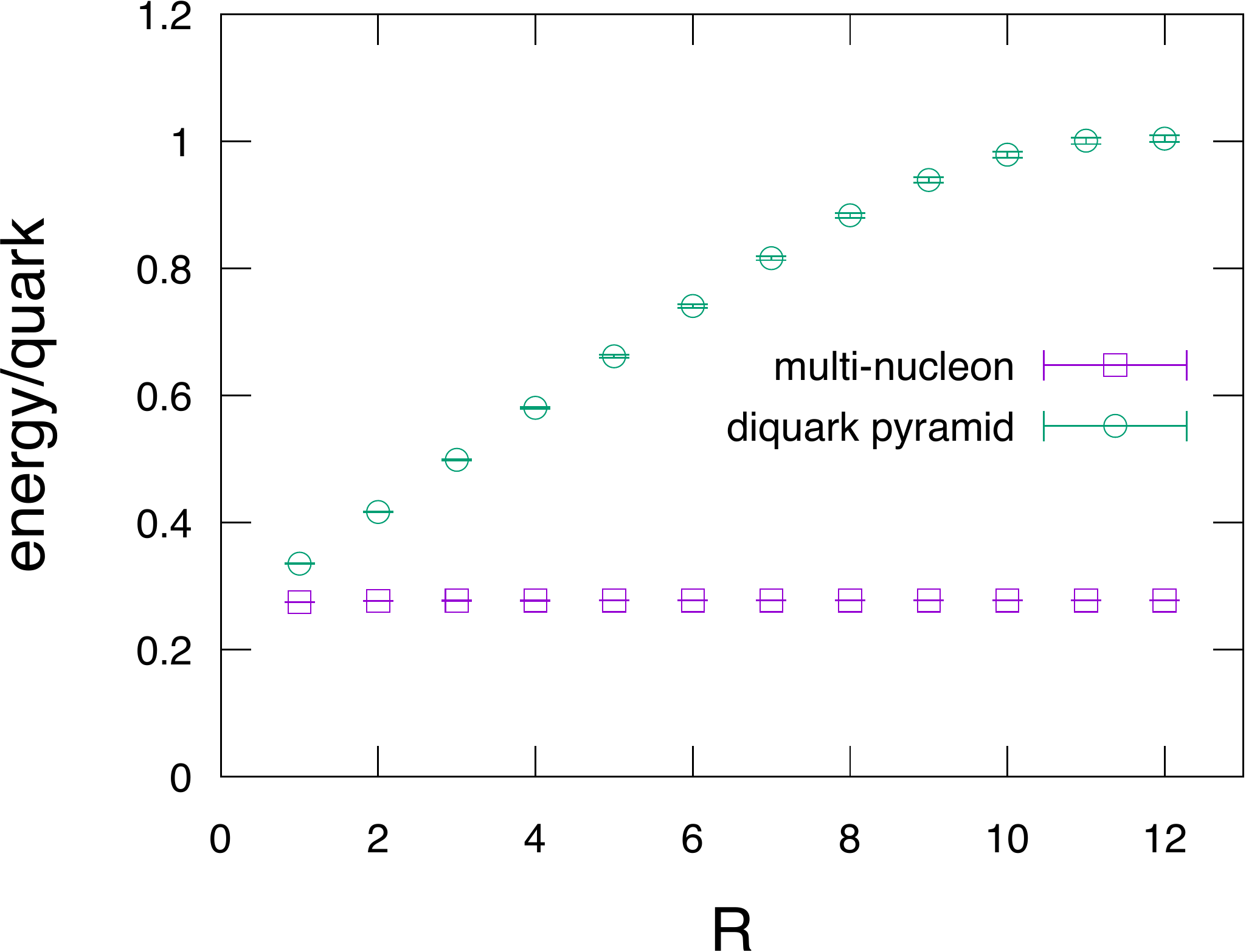}
}
\hfill
\subfigure[~ 12 quarks]  
{   
 \label{bhgas2}
 \includegraphics[scale=0.25]{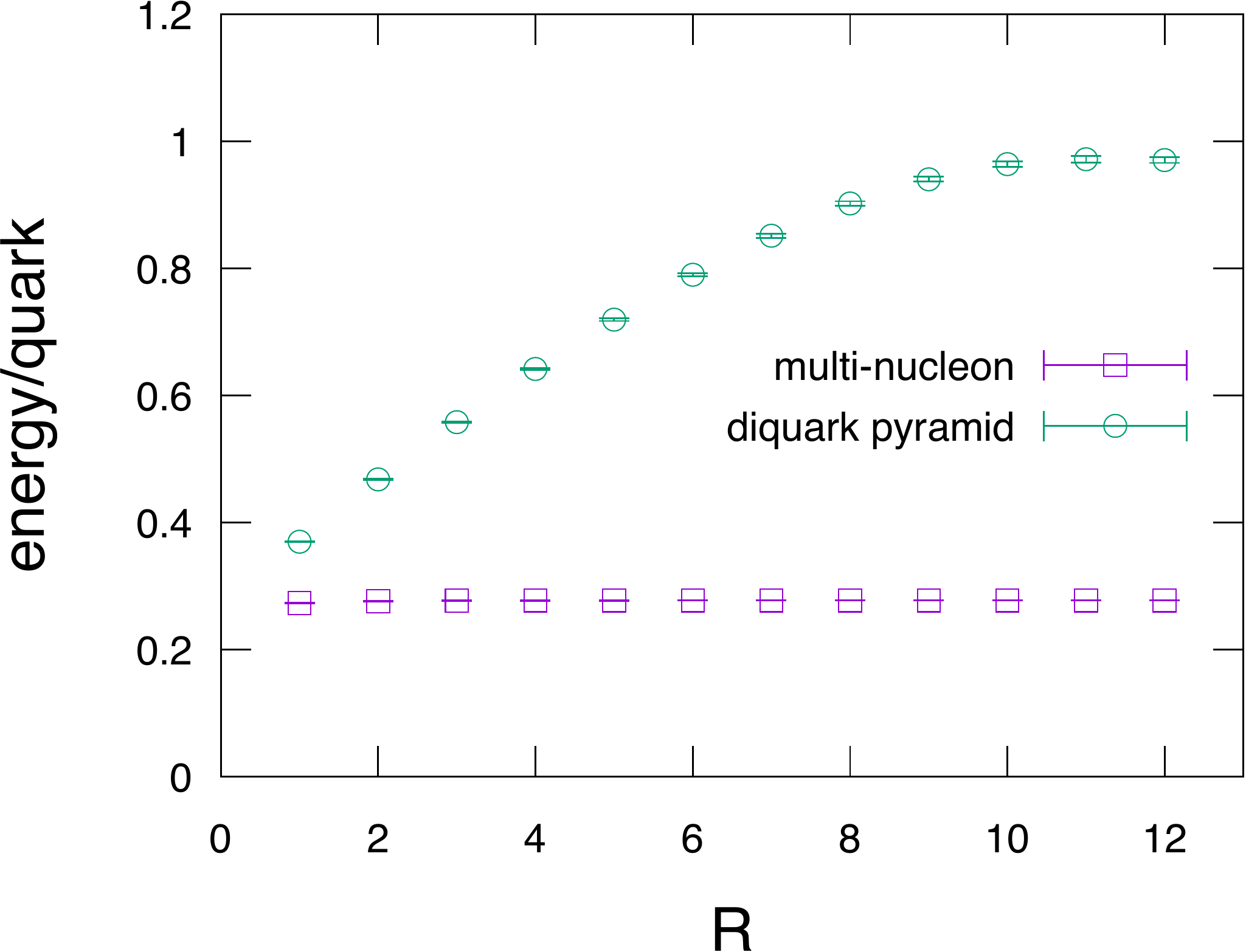}
}
\hfill
\subfigure[~ 24 quarks] 
{   
 \label{bhgas3}
 \includegraphics[scale=0.25]{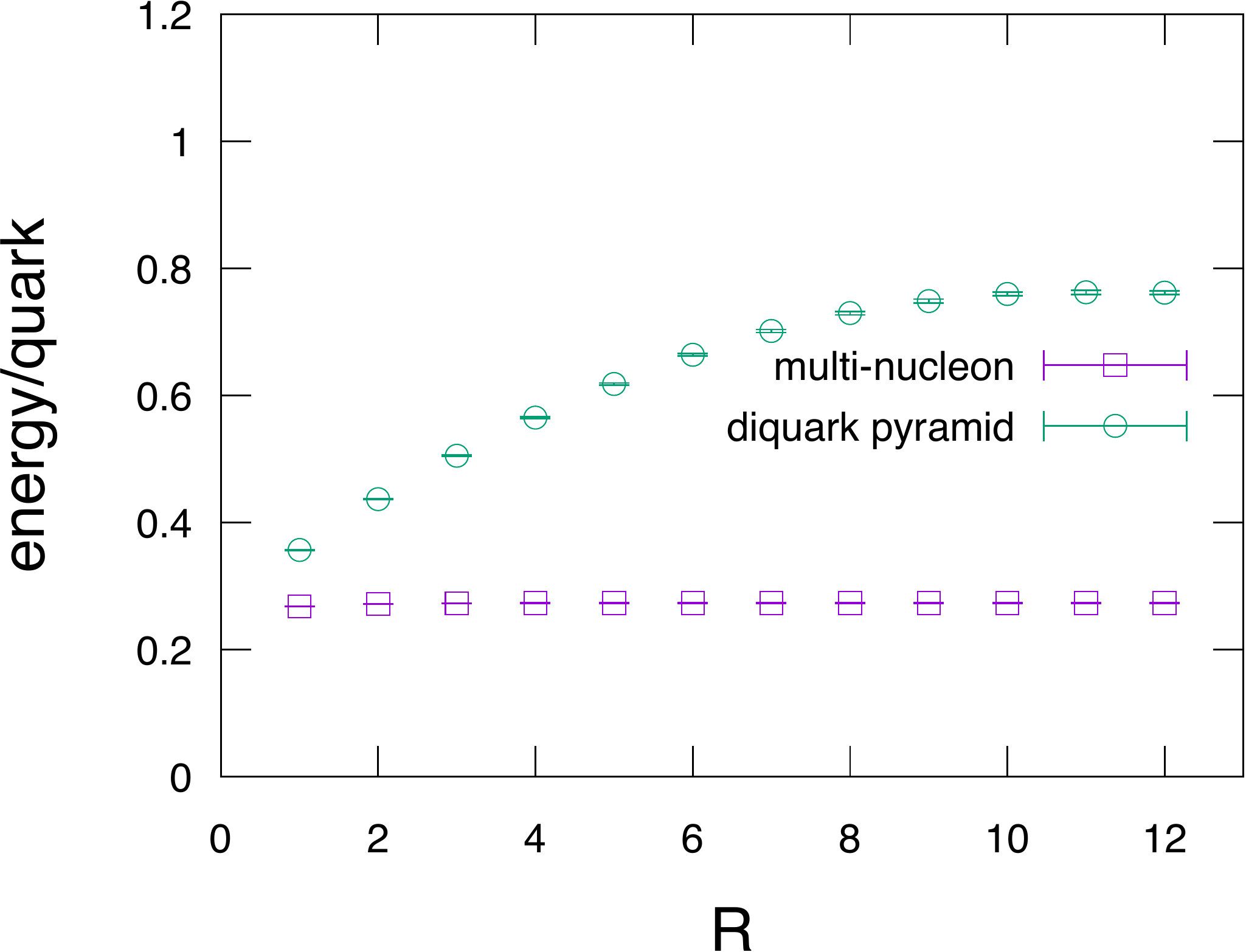}
}
\caption{Arrangements of 6, 12, and 24 static quarks (subfigures a-c).  In the multi-nucleon state, triplets of neighboring
quarks are contracted into singlets.  In the diquark pyramid the colors are contracted as specified in eqs.\ \rf{dqp6},
\rf{dqp12}, \rf{dqp24}, respectively.  The Coulomb energies per quark, as a function of the separation $R$, are displayed
for the 6, 12, 24 quark arrangements in subfigures d-f, respectively.}
\label{bhgas}
\end{figure*}
 
\section{Spatial arrangements and Coulomb energies} 
 
    We work in the framework of SU(3) lattice pure gauge theory with the usual Wilson action on a $24^4$ lattice volume, at lattice coupling $\b=5.8$.  
    
    Let us consider operators $L_a^{~~b}(i)$ and $M^a_{~~b}(i)$ which transform under the remnant global gauge symmetry
as $U_0(x)$ and $U_0^\dg(x)$ respectively, i.e.
\bea
           L_a^{~~b} &\ra& {L'}_a^{~~b} = g_a^{~~c} g^b_{~~d} L_c^{~~d} \non \\
           M^b_{~~a} &\ra& M'^b_{~~a}= g^b_{~~c} g_a^{~~d} M^c_{~~d} \ .
\eea
When contracting two quarks in the $\tbf$ representation into a diquark in the $\tbar$ representation, and allowing these charges to propagate for one lattice spacing in the time direction, then we are essentially contracting two $U_0$ operators into an $M$ operator.  Likewise, contracting two diquarks in the $\tbar$ representation into an operator in the $\tbf$ representation, and propagating for one lattice spacing in the time direction, involves contracting two $M$ operators into
an $L$ operator.  It will be helpful to introduce the following notation:
\begin{enumerate}
\item contraction of three $L$ or three $M$ operators to a singlet:
\bea
          [1:2:3] &\equiv& \left\{ \begin{array}{c}
      \e^{a_1 a_2 a_3} \e_{b_1 b_2 b_3} 
         L_{a_1}^{~~b_1}(1) L_{a_2}^{~~b_2}(2) L_{a_3}^{~~b_3}(3)  \cr \cr
      \e_{a_1 a_2 a_3} \e^{b_1 b_2 b_3} M^{a_1}_{~~b_1}(1) M^{a_2}_{~~b_2}(2) M^{a_3}_{~~b_3}(3)
                    \end{array} \right. \ . \non \\
\eea
\item contraction of two $L$ operators to an $M$ operator, or contraction of two $M$ operators to an $L$ operator
\bea
          \{2:3\} &\equiv& \left\{ \begin{array}{c} 
          \e^{a_1 a_2 a_3} \e_{b_1 b_2 b_3}  L_{a_2}^{~~b_2}(2) L_{a_3}^{~~b_3}(3)  \cr \cr
          \e_{a_1 a_2 a_3} \e^{b_1 b_2 b_3}  M^{a_2}_{~~b_2}(2) M^{a_3}_{~~b_3}(3)
           \end{array} \right. \ . \non \\
\eea
\end{enumerate}
     As a warm-up exercise, we consider six ($p=1$) quarks in a plane, arranged in two L-shaped arrangements (indicated by the solid lines) of three quarks, with the groups separated by a lattice distance $R$, as shown in Fig.\ \ref{q6}.  In an MN arrangement, the three quark indices in each L-shaped subgroup are contracted into a singlet, producing two nucleon states.
The energy of this arrangement is obtained, as explained previously by computing the expectation value of
the operator
\bea
     Q_{MN}(U) =  [U_0(1):U_0(2):U_0(3)] \times [U_0(4):U_0(5):U_0(6)] \ , \non \\
\label{mnq6}
\eea
with integers $1-6$ corresponding to the six quarks shown in Fig.\ \ref{q6}.  The energy per quark is then given by
\beq
     \E_q = - {1\over 6} \log\left[{\langle Q(U) \rangle \over Q(\mathbbm{1}) }\right] \ .
\label{Eq}
\eeq
In the DQP arrangement, we pair each quark in one subgroup with a quark in the other subgroup to form three diquarks in the $\tbar$ representation, and then contract these three diquarks into a singlet as follows:
\beq
Q_{DQP}(U) =  [\{U_0(1),U_0(4)\} : \{U_0(2),U_0(5)\} : \{ U_0(3),U_0(6)\}] \ .
\label{dqp6}
\eeq
We compute the Coulomb energies per quark of the MN and DQP states from \rf{Eq}, as described previously, and find the results shown in Fig.\ \ref{bhgas1}.  These are as expected.  In the MN case, the energies are almost independent of the separation $R$ between the L-shaped nucleons.   In the DQP case, the energy rises linearly with separation (until the data begins to flatten out due to the finite lattice volume) and is strongly disfavored  energetically compared to the MN configuration.  

   The next possibilities are 12 ($p=2$)  and 24 $p=3$ quarks, with L-shaped groups of three quarks placed in the arrangements shown in Figs.\  \ref{q12} and \ref{q24} respectively. Again the $Q_{MN}$ operator is constructed by contracting indices in each L-shaped group to form a singlet, and then taking the product.  The $Q_{DQP}$ operator is created
by first contracting pairs of quark charges to form diquarks in the $\tbar$ representation, then contracting pairs of $\tbar$ diquarks into $\tbf$ combinations, and so on until finally contracting the three remaining operators into singlets.  Shortening the notation further to let an integer $i$ denote the link variable $U_0(i)$ associated with the $i$-th static quark, the choice
of contractions is, for the 12 quark combinations
\bea
 Q_{DQP}(U) = [\{\{1,4\},\{7,10\}\} &:& \{\{2,5\},\{8,11\}\} \non \\
 &:& \{\{3,6\},\{9,12\}\}] \ ,
 \label{dqp12}
 \eea
 and for the 24 quark combination
 \begin{widetext}
 \bea
 Q_{DQP}(U) = [\{\{\{1,4\},\{7,10\}\},\{\{13,16\},\{19,22\}\}\} &:& \{\{\{2,5\},\{8,11\}\},\{\{14,17\},\{20,23\}\}\} \non \\
  &:& \{\{\{3,6\},\{9,12\}\},\{\{15,18\},\{21,24\}\}\}] \ .
 \label{dqp24}
 \eea
 \end{widetext}
 We compute energies $\E_q$ per quark from $-\log [\langle Q(U)\rangle /Q(\mathbbm{1}]$, and dividing by
 the number of quarks, either 12 or 24 in these cases.
The energies of the corresponding DQP and MN states for these
12 and 24 quark states vs.\ $R$ are shown in Figs.\ \ref{bhgas2} ($p=2$) and \ref{bhgas3} ($p=3$) respectively.  Again this is what one would expect; in both cases the MN state is independent of the separation $R$, while the energy of the DQP state rises linearly with $R$.

\begin{figure}[htb]
 \includegraphics[scale=0.3]{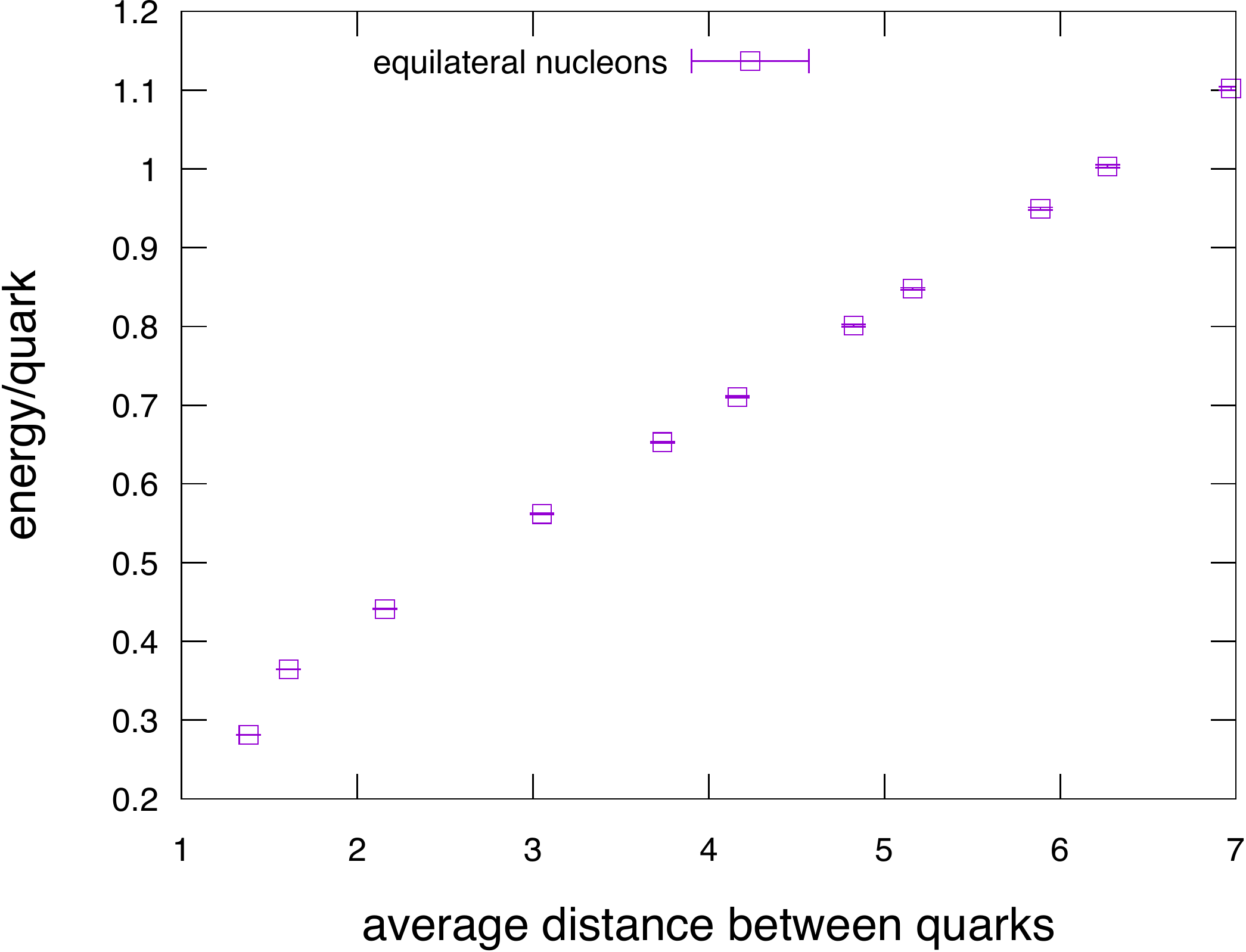}
\caption{The Coulomb energies of a nucleon state consisting of three static quarks in an arrangement as
close as possible (within the constraints of the lattice structure) to an equilateral triangle.  The figure displays the
energies of such states as a function of the average interquark separation in the state.} 
\label{triangle}
\end{figure} 

    Rather than keep three quarks in a nearest-neighbor L-shaped configuration, we may also compute the Coulomb
energy of a single three quark nucleon as a function of average quark separation within the nucleon.  In this calculation
we arrange the quarks to lie as closely as possible (within the constraints of the lattice structure) on an equilateral triangle, with the average quark separation defined as being 
$R = (d_{12}+d_{23}+d_{31})/3$, where $d_{ij}$ is the (straight-line) distance between quarks $i,j$ in the nucleon.
Once again, and unsurprisingly, the Coulomb energy per quark rises linearly with $R$, as seen in Fig.\ \ref{triangle}.

\section{Coulomb energy in a dense medium}

    Instead of keeping the number of quarks fixed and varying the separation between nucleons, as in the previous section, we next keep the density fixed at one quark per lattice site in a rectangular volume, and vary the size of the rectangular volume containing the quarks.  In this setup we choose to average over all possible color contractions of the MN and DQP type.  In other words, in the MN case, we group the quarks into $2^p$ randomly chosen sets of three, contracting the indices in each group into a singlet.  In the DQP scheme we group the quarks into $2^{p-1} \times 3$
randomly chosen sets of two quarks, contracting the indices to form a $\tbar$ combination.  These diquarks are then
randomly grouped into $2^{p-2} \times 3$ sets of two diquarks, contracting the indices in each set into a $\tbf$ combination,
and so on until only three $\tbar$ or $\tbf$ combinations are available, and these are finally grouped into a singlet.  
We then compute the Coulomb energy from correlation functions of timelike link variables, as explained above, and divide by the number of quarks to obtain the energy per
quark.   In this calculation the quarks are arranged in a rectangular volume of $i \times j \times k = 2^p \times 3$ lattice units, with each lattice site in the volume occupied by one quark.  At each
data-taking sweep through the lattice we compute the observable in each $ijk$ rectangular subvolume of the lattice, choosing a different random grouping in each subvolume.

   As an example, the expectation value of the $Q_{DQP}(U)$ operator, for the $3 \times 2 \times 4$ arrangement corresponding to Fig.\ \ref{q24} with $R=1$, involves taking the expectation value of the operator shown in \rf{dqp24},
but averaged over all permutations of the quark numbers shown.  Likewise, for the $3 \times 2 \times 1$ arrangement
corresponding to Fig.\ \ref{q6} at $R=1$, the energy of the multi-nucleon case is extracted from the expectation value of the operator  $Q_{MN}(U)$ as shown in \rf{mnq6}, but averaged over all permutations of the quarks in this expression.

\begin{figure*}[t!]
 \includegraphics[scale=0.45]{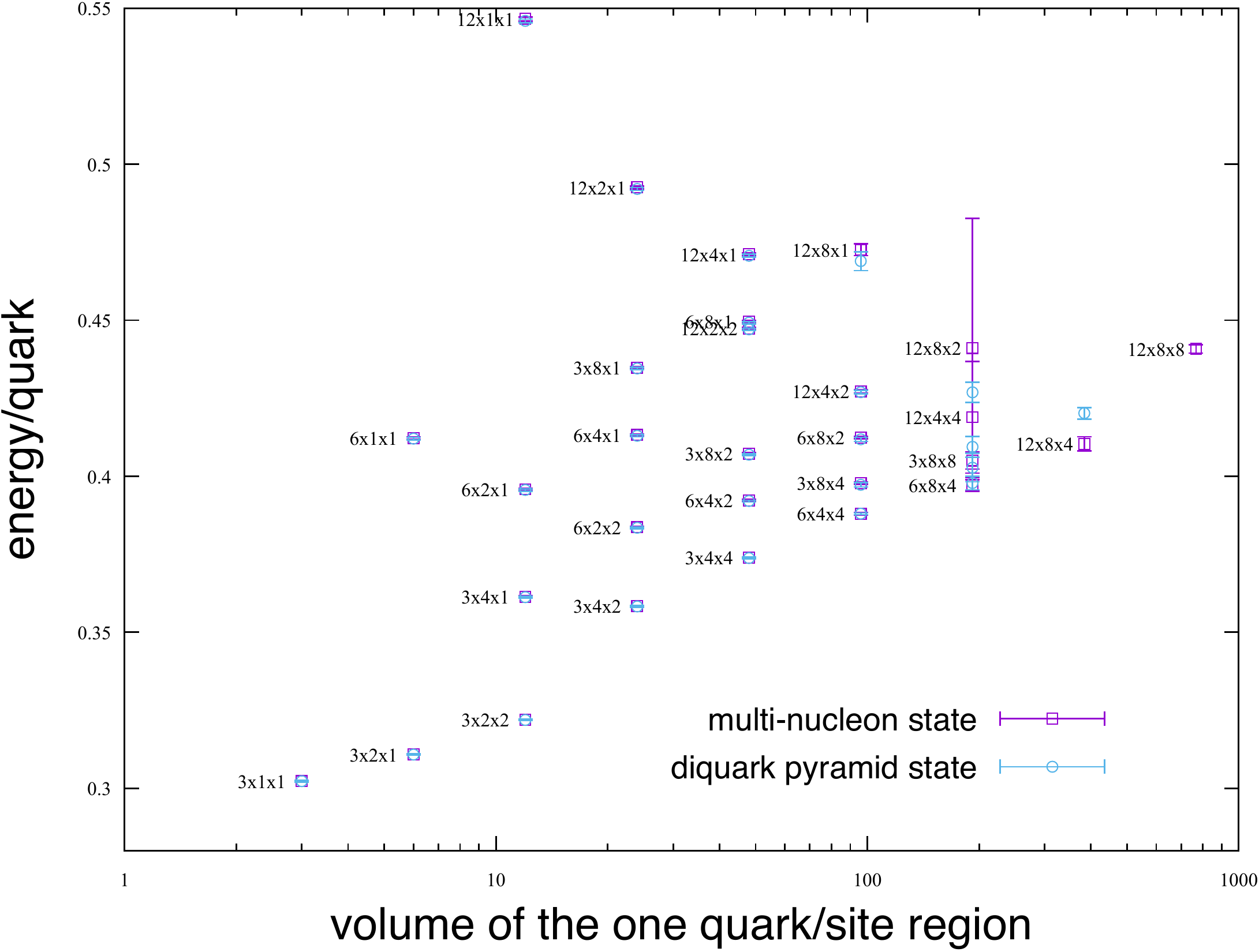}
\caption{Coulomb energies per quark for dense quark systems (one quark/site) in an $i\times j \times k$ volume, plotted
versus the volume.  The multi-nucleon
state groups the quarks at random into sets of three, contracting each triplet into a singlet.  The diquark pyramid 
contractions are also randomized, as described in the text.} 
\label{bh_random}
\end{figure*} 

\begin{figure*}[t!]
 \includegraphics[scale=0.45]{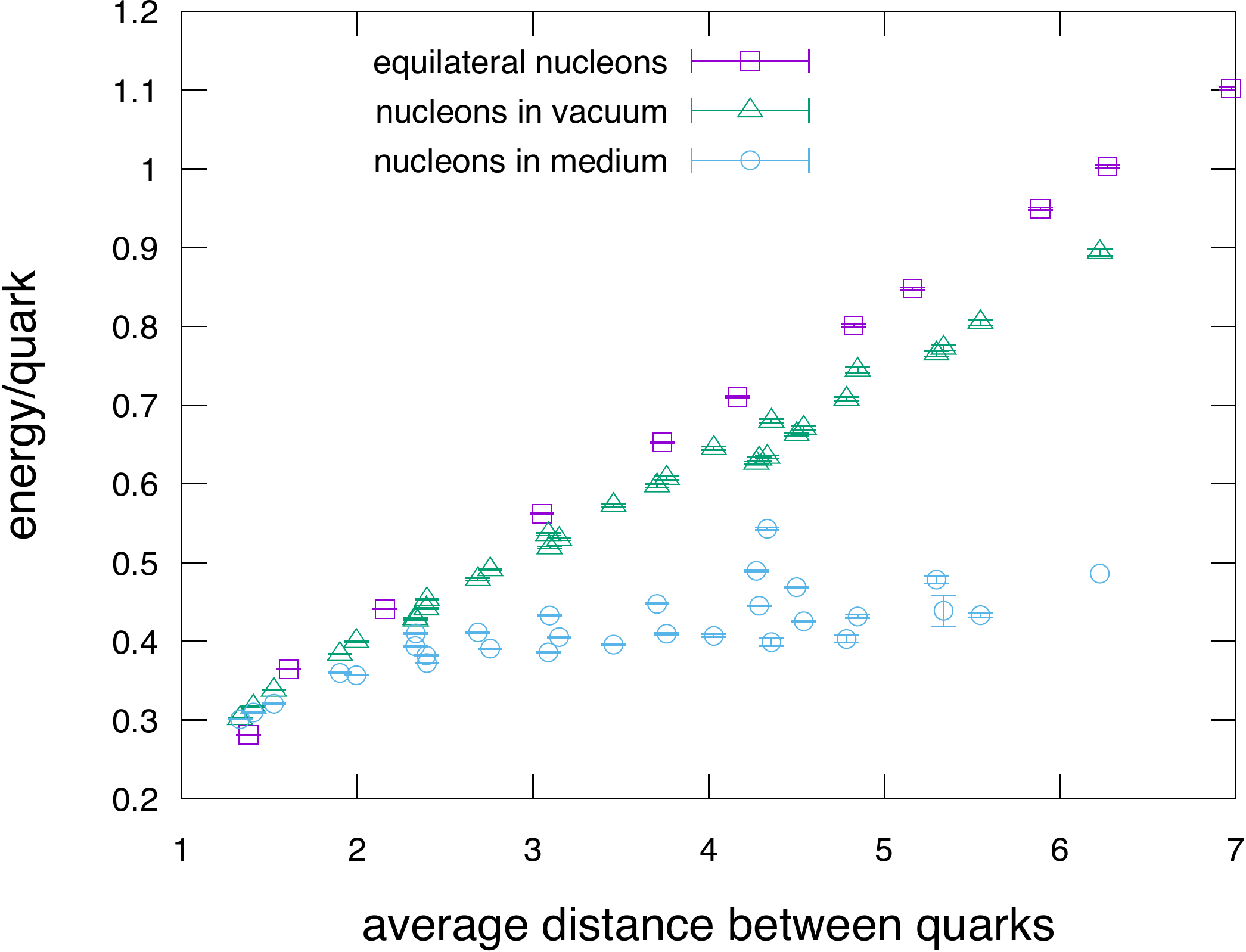}
\caption{The energy/quark in the multi-nucleon states (open circles) shown in the previous figure, plotted against the average separation between quarks in those nucleons (random triplets chosen at random in the $ijk$ volume, contracted to a singlet).  Also plotted are the energy per quark of isolated nucleons in vacuum (open squares) in a (nearly equilateral) arrangement, which was already plotted in Fig.\ \ref{triangle}, and of individual random nucleons (open triangles), selected as described in the text.} 
\label{bh_distance}
\end{figure*} 

      The results of this calculation, shown in Fig.\ \ref{bh_random} for each $ijk$ arrangement, are a little surprising.   Consider the MN contractions.  If the quarks are divided at random in groups of three, then the average separation between quarks in each group, denoted by $R$, ought to grow with the volume.  Yet the dependence of energy per quark 
 (= energy density $\E_q$)  turns out to be only mildly dependent on $i\times j\times k$ volume, even as we go from 6 quarks at $ijk=(6,1,1)$ and $R=2.33$, where we find $\E_q = 0.41$, to 768 quarks at $ijk=(12,8,8)$ and $R=6.23$ with $\E_q=0.44$.
In Fig.\ \ref{bh_distance} we plot $\E_q$ vs.\ $R$ for the MN states of quarks in the rectangular volume, as compared to the energy per quark vs.\ $R$  for three isolated quarks in a roughly equilateral arrangement, which was already displayed in Fig.\ \ref{triangle}.   We have also computed
 the energy of single nucleon states extracted from the $ijk$ volumes, denoted ``nucleons in vacuum'' in the figure.  Here we have computed the single nucleon energies by contractions $[a:b:c]$ for quarks $a,b,c$ chosen at random in the $ijk$ volumes, and averaged over the possible choices of $a,b,c$.  As opposed to the MN states we do not take the product
of such contractions, but only take the logarithm of the expectation value of single nucleon contractions to obtain a single nucleon energy, which 
we have again plotted against the average quark separation.  These energies represent the energies of nucleons in a vacuum, rather than in a medium, although not necessarily in an equilateral arrangement.  It can be seen that the energies of the ``nucleons in vacuum'' are roughly parallel, as a function of average quark separation, to the energies of quark triplets in
an equilateral configuration.
 
       What is clear from Fig.\ \ref{bh_distance} is that the average Coulomb energy per quark in a nucleon embedded in a dense medium is only mildly dependent on the average separation of the quarks within a nucleon, whereas the energy
clearly increases linearly with separation for an isolated triplet of quarks.  Since this is the central result of our work, it may be worth repeating:  in the MN case
we choose sets of three quarks in the $(ijk)$ volume at random, and of course this means that the average interquark
separation of quarks within each nucleon increases at the rectangular volume increases.  The surprise is that this
increase in separation in only very weakly reflected in the Coulomb energy per quark.  If the quarks were dynamical
this would not necessarily be a surprise; the phenomenon might be expected from a Debye screening process of
some kind.  But with static quarks, and an energy which derives from the instantaneous and non-local Coulomb term
in the Coulomb gauge Hamiltonian, it is hard to see how there can be any Debye-like screening process.  So this
result we consider a surprise.  We do not, at present, have an explanation for this effect.

      The energy/quark $\E_q$ for the random DQP arrangements described above are also displayed, together
with the previous MN results, in Fig.\ \ref{bh_random} .  What is again a little surprising, at least to us,  is that the MN and
DQP energies per quark are quite comparable.  The DQP arrangement of dense static quarks should be preferred, thermodynamically, just on considerations of multiplicity, as noted in section \ref{contra}.   

\begin{figure}[t!]
 \includegraphics[scale=0.35]{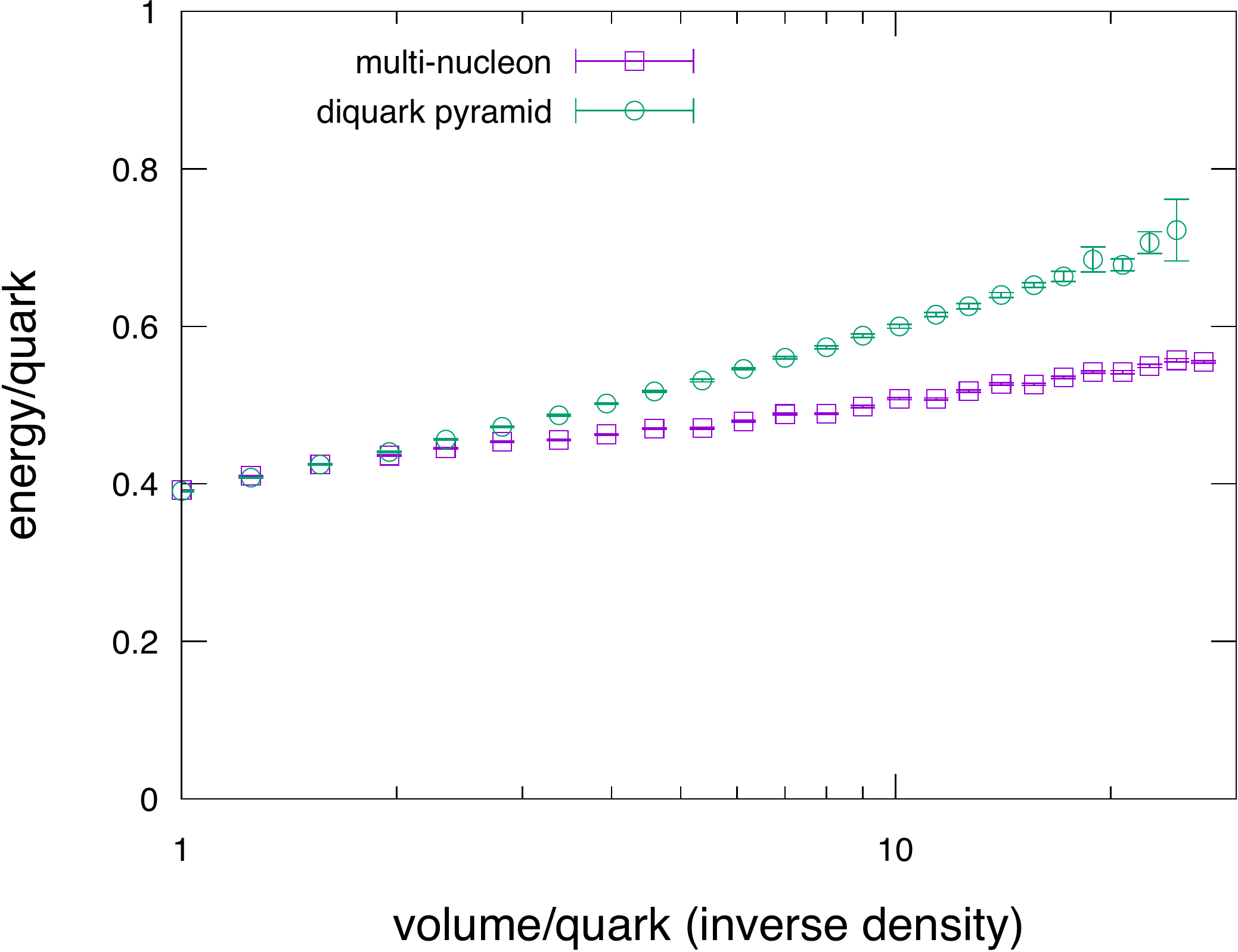}
\caption{The energy per quark of a system of 48 quarks in which the density of the system is varied by increasing
the volume.  The plot is energy per quark vs.\ inverse density.} 
\label{bh_density2}
\end{figure}

     Finally we have studied how $\E_q$ changes (in lattice units at $\b=5.8$) as the density of quarks in the $ijk$ volumes
are varied away from the density of one static quark per lattice site.  For this purpose we start with 48 quarks in an
$(ijk)=(3,4,4)$ arrangement.  Keeping the number of quarks fixed at 48, we increase the volume by adding, at each step,
one unit to the length of the volume in the $x,y,z$ directions, in that order.  In other words, we increase the volume 
in the following order: 
\begin{widetext}
\bea & &(i+1,j,k),(i+1,j+1,k),(i+1,j+1,k+1),(i+2,j+1,k+1),  \non \\
 & & \qquad \qquad (i+2,j+2,k+1),(i+2,j+2,k+2),(i+3,j+2,k+2),...
\eea
\end{widetext}
The 48 quarks are placed at random in each of the larger $(i',j',k')$ volumes, with the MN and DQP contractions
constructed as before.  
We observe in Fig.\ \ref{bh_density2} that the energy per quark does increase somewhat as the 
density decreases (volume/quark increases), as one might expect, but the increase in energy in the DQP and (especially)
the MN cases are comparatively modest, given an order-of-magnitude reduction in the quark density.

\begin{figure}[thb]
 \includegraphics[scale=0.35]{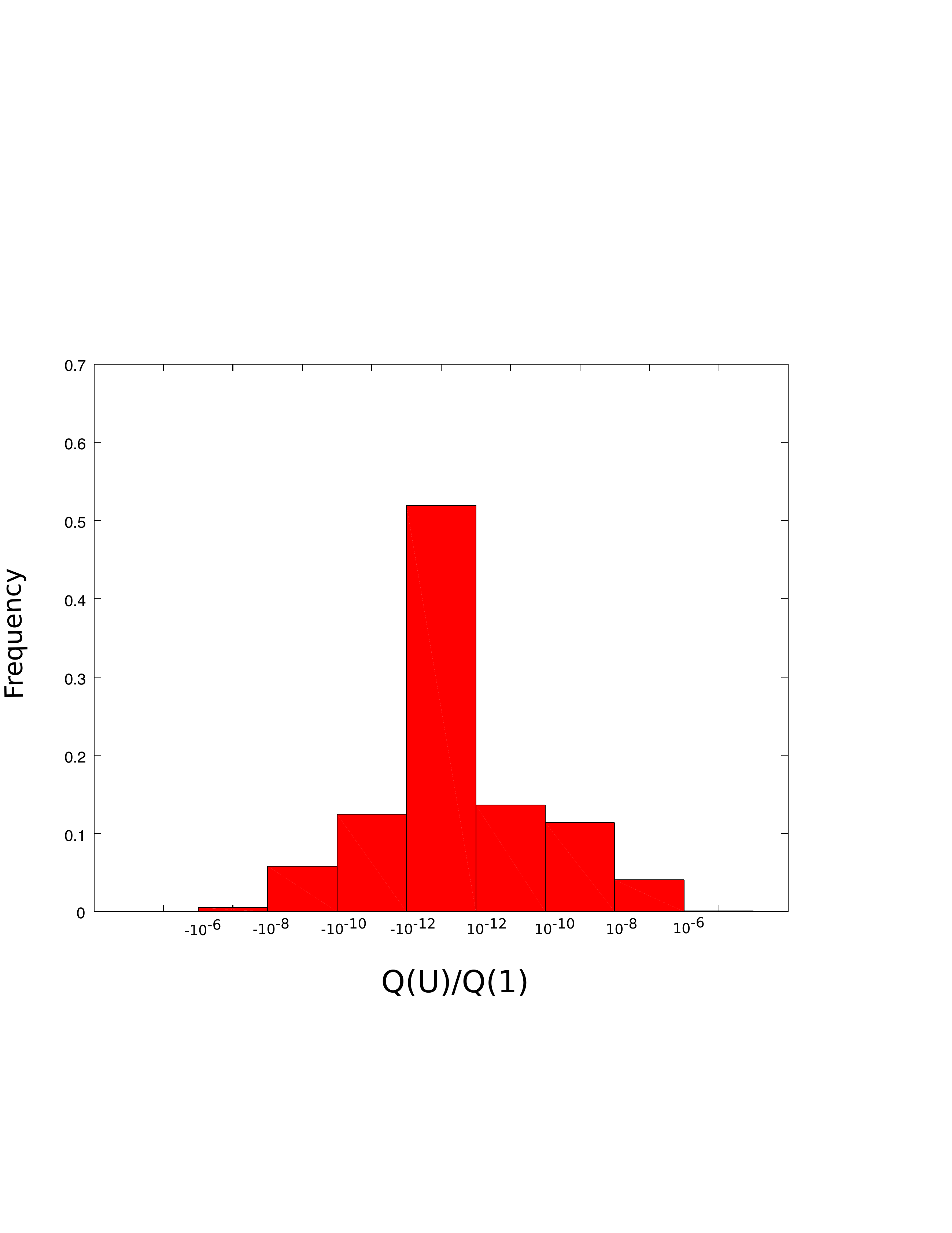}
\caption{Histogram of 100,000 values of $Q(U)/Q(\mathbbm{1})$ obtained for 48 static quarks in a $3 \times 4 \times 4$ volume. The central bin shows the frequency of measured values lying between $-10^{-12}$ and $+10^{-12}$, the adjacent bin to the right is the frequency of measured values in the range $10^{-12}$ to $10^{-10}$, and so on.  Note the small
magnitude of these values, and the asymmetry in the frequencies of positive and negative values.}
\label{histplot}
\end{figure}

      In the Introduction we mentioned the sign problem in connection with high baryon density, which makes a
conventional Monte Carlo calculation (e.g.\ via re-weighting) unfeasible due to sign cancellations.  Inspection of Fig.\ 
\ref{bh_random}, which displays on the $y$-axis the logarithm of $\langle Q(U)/Q(\mathbbm{1}) \rangle$ divided by
the number of of quarks, makes it clear that $\langle Q(U)/Q(\mathbbm{1}) \rangle$ itself must have extraordinarily small values for our larger volumes of quarks, and the question is how it is possible that such minute values are not swamped by statistical error.  The answer is that these very small values are not, for the most part, due to delicate sign cancellations among measured values of O(1), but rather come mainly from the very small magnitude of each measured value.  This is illustrated for the $3\times 4 \times 4$ volume by the histogram of $Q(U)/Q(\mathbbm{1})$ values shown in Fig.\ \ref{histplot}. The set consists of 100,000 values obtained at $\b=5.8$ on a $10^4$ lattice.  The average value in this case is $7.06 \times 10^{-9}$, and the asymmetry between positive and negative values, which results in this non-zero average, is obvious at a glance.
 
\section{Conclusions}

    Studies of the Coulomb energy in static quark systems have generated some unexpected results.
In the past it was found that the Coulomb energy in a static quark-antiquark pair
rises linearly with quark separation \cite{Greensite:2003xf,*Nakagawa:2006fk,Greensite:2014bua}, contains a L\"uscher term \cite{Greensite:2014bua}, and (here is the surprise) it is arranged into a flux tube which is somewhat more narrow than the minimal energy configuration \cite{Chung:2017ref}, rather than being spread out over all space as one might naively expect.
 
     We have now investigated the Coulomb energy of a system of $2^p\times 3$ static quarks, at a density of one
quark per lattice site in a rectangular volume, with colors contracted into either a set of $2^p$ ``nucleons,'' the MN state, or into a di-quark
``pyramid,'' the DQP state, which cannot be factorized into subsets of color singlets.  We average over the possible 
selections of three quarks into nucleon subsets,
or the choice of quark pairs contracted into diquarks in the DQP state, as explained previously.
What we have found is that the energy per quark, in either the MN or DQP states, is only very weakly dependent
on the rectangular volume containing the quarks.  The reason this was unexpected, especially in the MN case, is that
while the average separation between quarks in a nucleon rises with volume, this is not reflected in a corresponding
linear rise in the energy per quark.  This behavior of the energy of a nucleon in a dense medium contrasts sharply with
the corresponding energetics of an isolated nucleon with static quarks, where the energy per quark does rise linearly
with quark separation.

   In a plasma of dynamical quarks, this insensitivity of energy/quark to the volume of the ensemble of quarks would
not be especially surprising, and would be explained via some Debye-like screening mechanism in the quark
plasma.  But since we are dealing with an ensemble of {\it static} quarks, whose Coulomb interactions presumably arise from
the instantaneous non-local term in the Coulomb gauge Hamiltonian, it is difficult to appeal to such screening mechanisms.
We appear to be finding some unexpected property of the Coulomb energy of a dense ensemble of quark charges,
a kind of ``screening-without-screening,''  such that the long-range interactions of static quarks are somehow damped
without corresponding quark motion.   Our data suggests that the Coulomb energy density of a state at fixed baryon number density is fairly insensitive to the total number of quarks, regardless of how the color indices are contracted.
This is obviously a non-perturbative effect.  

   It should be understood that although the states we are constructing are physical states, whose interaction energy
is the Coulomb energy, these are not necessarily the minimal energy states of
a dense quark system.  For example, the analogous states of a quark-antiquark pair have a string tension, due to the
Coulomb interaction, which is about four times the asymptotic string tension.  Nevertheless, qualitative features of
such states, such as the linear potential, the existence of a flux tube, and the L\"uscher term, persist in the minimal
energy state.  It is possible that ``screening-without-screening'' is a feature of the minimal energy version of a
heavy dense quark ensemble, which may have implications also for light quarks at finite densities.  We leave this
possibility for future study.

\acknowledgments{This work is supported by the U.S.\ Department of Energy under Grant No.\  DE-SC0013682.}   

\bibliography{obh}

\begin{thebibliography}{1}

\bibitem{Aarts:2016qrv}
G.~Aarts, F.~Attanasio, B.~Jager, and D.~Sexty,
\newblock JHEP {\bf 09}, 087 (2016), arXiv:1606.05561.

\bibitem{Langelage:2014vpa}
J.~Langelage, M.~Neuman, and O.~Philipsen,
\newblock JHEP {\bf 09}, 131 (2014), arXiv:1403.4162.

\bibitem{Greensite:2003xf}
J.~Greensite and S.~Olejnik,
\newblock Phys.Rev. {\bf D67}, 094503 (2003), arXiv:hep-lat/0302018.

\bibitem{Nakagawa:2006fk}
Y.~Nakagawa, A.~Nakamura, T.~Saito, H.~Toki, and D.~Zwanziger,
\newblock Phys.Rev. {\bf D73}, 094504 (2006), arXiv:hep-lat/0603010.

\bibitem{Greensite:2014bua}
J.~Greensite and A.~P. Szczepaniak,
\newblock Phys.Rev. {\bf D91}, 034503 (2015), arXiv:1410.3525.

\bibitem{Anselmino:1992vg}
M.~Anselmino, E.~Predazzi, S.~Ekelin, S.~Fredriksson, and D.~B. Lichtenberg,
\newblock Rev. Mod. Phys. {\bf 65}, 1199 (1993).

\bibitem{Lichtenberg:1982jp}
D.~B. Lichtenberg, W.~Namgung, E.~Predazzi, and J.~G. Wills,
\newblock Phys. Rev. Lett. {\bf 48}, 1653 (1982).

\bibitem{Chung:2017ref}
K.~Chung and J.~Greensite,
\newblock Phys. Rev. {\bf D96}, 034512 (2017), arXiv:1704.08995.

\end{thebibliography}

\end{document}